\documentclass[letterpaper]{article} % DO NOT CHANGE THIS
\usepackage{aaai25}  % DO NOT CHANGE THIS
\usepackage{times}  % DO NOT CHANGE THIS
\usepackage{helvet}  % DO NOT CHANGE THIS
\usepackage{courier}  % DO NOT CHANGE THIS
\usepackage[hyphens]{url}  % DO NOT CHANGE THIS
\usepackage{graphicx} % DO NOT CHANGE THIS
\usepackage{natbib}  % DO NOT CHANGE THIS AND DO NOT ADD ANY OPTIONS TO IT
\usepackage{caption} % DO NOT CHANGE THIS AND DO NOT ADD ANY OPTIONS TO IT
\usepackage{algorithm}
\usepackage{algorithmic}

\usepackage{newfloat}
\usepackage{listings}
\usepackage{dcolumn}
\DeclareCaptionStyle{ruled}{labelfont=normalfont,labelsep=colon,strut=off} % DO NOT CHANGE THIS
\floatstyle{ruled}
\newfloat{listing}{tb}{lst}{}
\floatname{listing}{Listing}
\nocopyright %-- Your paper will not be published if you use this command
\usepackage[table, x11names]{xcolor}
\usepackage{booktabs}
\usepackage{multirow}

\title{Scenarios in Computing Research: A Systematic Review of the Use of Scenario Methods for Exploring the Future of Computing Technologies in Society}

\author{
    Julia Barnett\textsuperscript{\rm 1},
    Kimon Kieslich\textsuperscript{\rm 2},
    Jasmine Sinchai\textsuperscript{\rm 1},
    Nicholas Diakopoulos\textsuperscript{\rm 1}
}
\affiliations{
    \textsuperscript{\rm 1}Northwestern University,\\ \textsuperscript{\rm 2}Institute for Information Law, University of Amsterdam\\
    juliabarnett@u.northwestern.edu
}

\begin{document}

\maketitle

\begin{abstract}
Scenario building is an established method to anticipate the future of emerging technologies. Its primary goal is to use narratives to map future trajectories of technology development and sociotechnical adoption. Following this process, risks and benefits can be identified early on, and strategies can be developed that strive for desirable futures. In recent years, computer science has adopted this method and applied it to various technologies, including Artificial Intelligence (AI). Because computing technologies play such an important role in shaping modern societies, it is worth exploring how scenarios are being used as an anticipatory tool in the field---and what possible traditional uses of scenarios are not yet covered but have the potential to enrich the field. We address this gap by conducting a systematic literature review on the use of scenario building methods in computer science over the last decade ($n=59$). We guide the review along two main questions. First, we aim to uncover how scenarios are used in computing literature, focusing especially on the rationale for why scenarios are used. Second, in following the potential of scenario building to enhance inclusivity in research, we dive deeper into the participatory element of the existing scenario building literature in computer science. 
\end{abstract}


\section{Introduction}

The use of narrativized scenarios in research has been a longstanding method dating back to the 1950s with early work focusing on military planning, public policy analysis, and strategic management decisions \cite{amer_review_2013, ramirez_plausibility_2014, ramirez2016strategic,van_der_heijden_scenarios_2005}. This method has gone by many names, spanning speculative design, what-if scenarios, scenario building, and more. Regardless of the terminology, the central thread of using scenarios is to transform complex information into a short story-like version that conveys essential information in an easy to understand manner to enable various stakeholders (experts or laypersons) to make informed opinions about possible futures. As Carroll puts it, ``Scenarios are not the traditional activity logs of human factors ... they are meaningful to and discussable by users; they are couched at the level at which people understand and experience their own behavior'' \cite{carroll1995scenario}.

%Recently, the use of this method has grown in popularity in the computer science domain, especially in regard to understanding the potential impacts of emerging technologies.
Due to the fragmented nature of scenario methods across various domains---there is no unified terminology for the method---it can prove difficult to understand the landscape of objectives researchers seek to achieve through the use of scenarios. For instance, researchers may get siloed in one body of work such as speculative design, and miss out on other areas like scenario planning and scenario building---which may result in blind spots and missing out on the potential for a more holistic application of scenarios. The use of scenarios in research can also enable a participatory element to engage with lay-stakeholders by making complex technology and its impacts more tangible and understandable. This paper thus offers (1) a springboard for those who already use scenarios to understand different objectives they can achieve, and (2) an introduction to scenarios and their potential for those who have never used them before. 

To understand this collective body of work within computing literature specifically, we have conducted a systematic literature review of the use of scenario methods in computer science over the last decade. We reviewed all research papers published in the Association for Computing Machinery (ACM) digital library from January 1 2015-January 31 2025 that use scenarios to perform futuring for any emerging technology. In particular, this paper explores the objectives computer science researchers are pursuing through the use of scenarios, as well as the degree of participation this method has enabled in the field of computing.

Our formal research questions are:

\begin{itemize}
    \item \textbf{RQ1}: What are the objectives of scenario building as described in computing literature, and what form are these scenarios taking?
    \item \textbf{RQ2}: How strong is the participatory element in scenario building studies, and how are researchers using scenario building to achieve this goal of participatory methods?
    % \item RQ3: Which terminology is used for scenario building, and are their different objectives and elements present when different terms are used?
    % \item RQ4: How are researchers generating these scenarios, and for whom are they writing them?
\end{itemize}

This paper makes two concrete contributions. First, it identifies five main ways that researchers in computing are using scenarios in their work: (1) to gather stakeholder needs and values, (2) to empower marginalized groups to imagine technology futures, (3) to provoke ethical reflection and promote critical awareness, (4) to anticipate threats and risks of these technologies, and (5) to explore perceptions and impacts of novel technologies before they launch. Second, it explores and quantifies the degree of participation this method can and has enabled in the computing literature.

% \begin{enumerate}
%     \item Contextualize what's going on: scenario building has been around for years across multiple domains and with multiple names. Recently, it has grown in popularity in the computer science field.
%     \item Highlight the issue: As a methodological tool, it has been used for a variety of goals, technologies, and participatory methods. When people working in this area go to cite related work, they often get siloed by one terminology (e.g., speculative design), and miss out on entire other areas. This work is fragmented and has many names, and thus makes it difficult to understand it as a collective body of work.
%     \item What are we doing about it: we perform a systematic literature review of all research papers published in the ACM digital library over the last 10 years that use scenario building to perform futuring for any emerging technology (ET) in order to capture how and why the CS field is using scenario building. 

% \end{enumerate}

\section{Background}

\subsection{The Use of Scenarios in Research}\label{sec:lr-scenario-building}

A scenario can be understood as a description of a potential future situation that traces back to current events, trends, and policies \cite{amer_review_2013}. 
%borjeson_scenario_2006
Scenario building does not need to describe \textit{one} state but opens the possibility to explore different future alternatives \cite{borjeson_scenario_2006}. We define \textbf{\textit{scenario building}} as a method that utilizes scenarios (typically in natural language text, but also in other media such as speech or video) that seeks to describe potential and alternative future outcomes and realities. Scenario building is often used as a strategic tool to reflect on potential strategies that might need to be taken if certain events take place (e.g., military strategies), or for business in respect to changing governance structures, business decisions, or changing user patterns \cite{selin_trust_2006}. Consequently, scenario building has had a long research tradition in the academic fields of economics and policy research, but also for research focusing on emerging technologies that has even resulted in the establishment of distinct academic journals like \textit{Futures} and \textit{Journal of Forecasting}. In addition to its use in scholarship, scenario building has also achieved broad adoption in companies %with the most prominent use as a strategic tool within the Royal Dutch/Shell Group 
\cite{varum_directions_2010}.
%amer_review_2013, 

Scenarios let users weigh multiple environmental impacts at once and reflect on the complexity of the world \cite{amer_review_2013}. They do not predict the future; they help users ``recognize, consider and reflect on the uncertainties they are likely to face'' \cite{varum_directions_2010}. Reducing uncertainty helps decision-makers prepare for and adapt to different situations \cite{nanayakkara_anticipatory_2020, uruena_understanding_2019}. 
%amer_review_2013, ramirez_plausibility_2014
Scholars emphasize synthesizing key elements into plausible, internally consistent future scenarios \cite{burnam-fink_creating_2015, ramirez_plausibility_2014}.
%amer_review_2013, selin_trust_2006
Scenarios are not neutral: ``Despite the aim to create objective or equally plausible scenarios, specific future worlds are sketched out and in the end scenarios are always selective and political'' \cite{selin_trust_2006}. They can amplify business strategies or highlight the voices of marginalized groups, invoking a participatory approach to sense-making. These participatory methods can emerge in \textit{participatory foresight} practices \cite{nikolova2014rise}.

Scenario building encompasses a wide variety of approaches. For instance, Börjeson et al. distinguish between \textit{predictive}, \textit{explorative}, and \textit{normative} scenarios, framed by ``What will happen?, What can happen? and How can a specific target be reached?'' \cite{borjeson_scenario_2006}. Each type favors distinct methods---from quantitative predictive models to qualitative, value-driven explorations of desirable or undesirable futures \cite{ amer_review_2013}. Scenario building can be achieved by the means of surveys, workshops, Delphi methods, etc., which also result in a large variance in the number of scenarios created and the time-frame of the scenarios \cite{borjeson_scenario_2006}. Under the umbrella term ``scenario building,'' different sub-categories and methods have evolved, including but not limited to \textit{science fiction prototyping }\cite{bray_radical_2022, burnam-fink_creating_2015}, \textit{storytelling} \cite{rasmussen_narrative_2005}, and \textit{speculative design} \cite{bray2021speculative, hohendanner2024metaverse}.%, and even as a method to develop \textit{forecasts} or \textit{foresight} \cite{amer_review_2013, brey_sienna_2022, floridi_ethical_2020}.

With their focus to illuminate plausible future developments, scenarios are especially useful to uncover trajectories of technology development. There has been scholarly work on using scenario building as an anticipatory tool for emerging technologies, especially those that might have severe consequences \cite{diakopoulos_anticipating_2021, mittelstadt_how_2015}. Brey states, ``The exploration of possible and plausible futures may provide valuable information. First, by giving glimpses of what may happen, it allows for better anticipation of the future than would be possible in a situation in which one has no idea what may happen. Second, by projecting possible future applications and uses of the technology and resulting consequences, it is possible to identify potential risks and benefits'' \cite{brey_ethics_2017}. 

Many of the most influential developments in technology in recent years---%can be traced back to computing technologies %and all the technological artifacts associated with them, be it 
from virtual reality tools to AI---stem from computing. In order to anticipate the impact of these technologies, scenario methods have entered computing disciplines, and scholars in this discipline have begun to adapt this research method to fit their objectives. Research disciplines shape the adoption of methodological toolkits to their modes of practice, and %it is therefore likely that scenario methods will be modified to fit disciplinary traditions and the technological artifacts at hand. 
computing will inevitably reshape scenario methods to fit its technological artifacts. Given computing’s societal reach, it is worth exploring how scenarios are being used as an anticipatory tool in the field---and what possible traditional uses of scenarios are not yet covered but have the potential to enrich the field. Yet, to our knowledge, there have only been literature reviews on the scenario building literature in the field of future studies \cite{amer_review_2013, borjeson_scenario_2006, varum_directions_2010}. This paper closes that gap with a review of scenarios in the field of computing. 

\subsection{Participatory Methods and Anticipatory Governance in Computing}

Understanding the vast assortment of risks and challenges from emerging technologies has traditionally been an expert-heavy endeavor \cite{herdel2024exploregen, solaiman2023evaluating, weidinger2021ethical}, which have been shown to be prone to biases \cite{bonaccorsi2020expert, brey_ethics_2017}. However, there has been a recent call to attention for the need of more inclusive methods to involve a more diverse variety of stakeholders \cite{kieslich2025scenario, metcalf2021algorithmic}---whether laypeople or ``other experts'' from adjacent domains \cite{nikolova2014rise}. This is especially important when considering the disciplinary, political, and societal goal is to design emerging technologies in alignment with ``agreed upon'' morals of society, which are ever-changing and differ vastly when you query different stakeholders \cite{baum2020social}.

Participatory methods also involve including diverse voices who may otherwise not have the chance to contribute to the trajectory of a project, initiative, or other endeavor \cite{delgado_participatory_2023}. \citet{sclove2010reinventing} specifically defines participatory technology assessment as enabling ``people who are otherwise minimally represented ... to develop and express informed judgments concerning complex topics.'' Scenario building is particularly apt at assisting laypeople with the development of informed judgments concerning complex topics since they are able to interact with tangible narratives around technologies, or even transform their own ideas into short stories or excerpts. 

Beyond any moral arguments, participatory methods have been described as an approach that can improve the quality of future simulations by making them more realistic for a wider array of people \cite{barreteau2013participatory}. Participatory foresight in particular---a technique to anticipate future outcomes---proposes including diverse stakeholders (whether different types of ``experts'' or laypeople) to provide a more realistic and socially grounded image of reality and impact \cite{nikolova2014rise}.

\citet{delgado_participatory_2023} outline four levels of participation with stakeholders in participatory designs: (1) consult (2) include (3) collaborate, and  (4) own. Various methods can satisfy different levels, and scenario building can fulfill all four levels, depending on how it is used. %Scenario building is only one method that has been deployed in computer science literature to anticipate potential impacts of new technologies, and they all fall on different areas of participatory methods. 
Other expert-heavy methods %that contain little to no degree of participation 
including red teaming \cite{feffer2024red} and algorithmic audits \cite{bandy2021problematic} have gained popularity specifically with generative AI models.%, though they notably have a tendency to be mostly (if not exclusively) expert driven since they rely on technical expertise to carry out the methods. 
Various methods have emerged for the \textit{design} of new technologies that integrate participatory approaches which can tackle some overlapping aspects of impact assessment early on, such as user-centered design \cite{council2005double} or participatory design \cite{gregory2003scandinavian, simonsen2013routledge}. \citet{andersen_stakeholder_2021} note various methods that integrate stakeholder input and their respective drawbacks: workshops are time consuming and have asymmetric engagement, and interviews have limited knowledge sharing among participants. Scenario building can be used to complement and offset the drawbacks of various other participatory methods. It can be deployed in a quick and efficient manner for participants if needed, take the form of a collaborative group discussion if preferred, or even be integrated to workshops, interviews, or a survey-based structure. 

%\subsection{Definitions and Research Questions}

% This work focuses on scenario building in computing literature that explores the potential future of emerging technologies. Leaning on the definitions and uses described earlier in Section \ref{sec:lr-scenario-building} we define: 

% \begin{itemize}
%     \item \textbf{\textit{Scenario building}} as a method that utilizes scenarios (typically in natural language text, but also in other media such as speech or video) that seeks to describe potential and alternative future outcomes and realities.  
    % \item \textbf{\textit{Emerging technologies}} according to Rtolo et. al's understanding \cite{rotolo2015emerging} as technologies possessing the key attributes ``(i) radical novelty, (ii) relatively fast growth, (iii) coherence, (iv) prominent impact, and (v) uncertainty and ambiguity.''
% \end{itemize} With these definitions in mind, we focus on the following research questions:

%Yet, computer science as a research field is especially interesting to focus on as scholars posses a high amount of technical knowledge and/or are even building the technical artifacts themselves. On the other hand, as scenario planning is not an inherent approach in the field, scholars need to adopt and adapt the approach for their purposes which might lead to distinct and unique questions and approaches. Thus, this paper aims to review the literature on scenario planning in the computer science literature. Specifically, we focus on several research questions:

\section{Data and Methodology}

In order to comprehensively examine these research questions, we conducted a systematic literature review (SLR) of full research articles published in the ACM digital library over the last ten years. In order to be transparent and replicable, the reporting of this SLR was guided by the standards of the Preferred Reporting Items for Systematic Reviews and Meta-Analyses (PRISMA) guidelines \cite{page2021prisma}. We first established search criteria, then conducted an abstract screening phase, and then assessed the full text of the articles for inclusion eligibility. Our final coding documents are available at: \url{https://tinyurl.com/scenario-slr}.

\subsection{Search Strategy}

\subsubsection{Inclusion and Exclusion Criteria}\label{sec:inclusion_criteria}

\textbf{\textit{Formatively}} we analyzed full research articles published in English in the ACM digital library that utilized scenario building in their methods or analysis. We excluded extended abstracts, book chapters, write-ups proposing workshops or panels, short papers, and any other non-full-length research article. We analyzed the full contents of these works including supplemental material in the main PDFs. We did not consult outside material such as websites with accompanying appendices.

\textbf{\textit{Temporally}} we analyzed works that had been published in the last ten years. %We initially focused our study on scenario building for AI technologies, but wanted to expand our time range to the last year to understand this method has been applied to various emerging technologies in recent years. 
We decided to scope the time frame to ten years in order to track the latest developments in computing technologies---including the recent rise of generative AI---but also give space for other technologies that emerged during this time span. %That gave us both the opportunity to engage with the newest developments by maintaining an open focus on various computing technologies. 
For the context of this work, that meant they were published between January 1, 2015 and January 31, 2025. This covers a comprehensive snapshot of how computer scientists are currently utilizing scenario building, as well as provides some historical context for how use of the method has morphed over the last decade. 

\textbf{\textit{Topically}} these papers had to be about using scenario building to explore future possible developments or impacts of emerging technologies. This excluded works that were using scenario building for non-technology futuring, such as exploring future ways to design museum exhibits. It also excluded works conducting scoping analyses or other SLRs (one explored speculative design in sustainable HCI \cite{soden2021we}, and the other explored 11 works that automatically constructed scenarios \cite{davis2023towards}). We lean on the definition proposed by \citet{rotolo2015emerging} of an \textbf{\textit{emerging technology}}, which identifies the key attributes as ``(i) radical novelty, (ii) relatively fast growth, (iii) coherence, (iv) prominent impact, and (v) uncertainty and ambiguity.'' We are liberal in our application of this definition (erring on the side of inclusion), for instance we consider any new technology warranting the writing of a full research paper to qualify as ``prominent impact.'' We found the description of radical novelty quite useful, however, as it allowed us to exclude descriptions of ``new'' technologies that had mostly matured or were simply being applied in a new context. We also exclude technologies that were entirely based on fiction (as of Spring 2025) such as technology manipulating time \cite{behzad2023co} or dreams \cite{bonlykke2024taking}. The emphasis on uncertainty and ambiguity aligns well with the goal of scenario building for uncovering potential impacts and developments of technologies. %Finally, during the article eligibility phase we exclude speculative design works that focus on improving specific design features or uses of technologies rather than focusing on future impacts of the technology on communities or society at large, such as works that conduct interviews with their potential users to understand how they would like to see prototypes altered. These works were excluded because they are not exploring potential futures of technology, but rather trying to improve one specific prototype for a userbase. 

\subsubsection{Keyword Search}
Based on this inclusion and exclusion criteria, we arrived at a final list of keywords with which to query the ACM digital library. We start with the main list of terms we are familiar with to describe scenario building, such as ``scenario writing'' and ``speculative fiction,'' and then we bolster our list with those identified by \citet{amer_review_2013} in their review of scenario planning. The finalized set of query terms for the ACM digital library was:

\begin{itemize}
    \item (Abstract OR Title contains: ``scenario*building'' OR ``scenario*writing'' OR ``scenario*based'' OR ``scenario*driven'' OR ``speculative*fiction'' OR ``speculative*design'' OR ``predictive*forecast'' OR ``what-if*scenario'' OR ``explorative*scenario'' OR ``strategic*scenario'' OR ``normative*scenario'' OR ``transforming*scenario'') AND
    \item E-Publication Date: (01/01/2015 TO 01/31/2025)
\end{itemize}

We noticed that the most recent proceedings of AI, Ethics, and Society (AIES '24) were not present in the ACM digital library yet part of the the broader ACM catalogue even though past years were, so we also manually queried the AIES proceedings page\footnote{https://ojs.aaai.org/index.php/AIES/search/search} with the same set of keywords to identify four more records to include in our search. 

\subsection{Title and Abstract Screening}

\begin{figure}
    \centering
    \includegraphics[width=.95\linewidth]{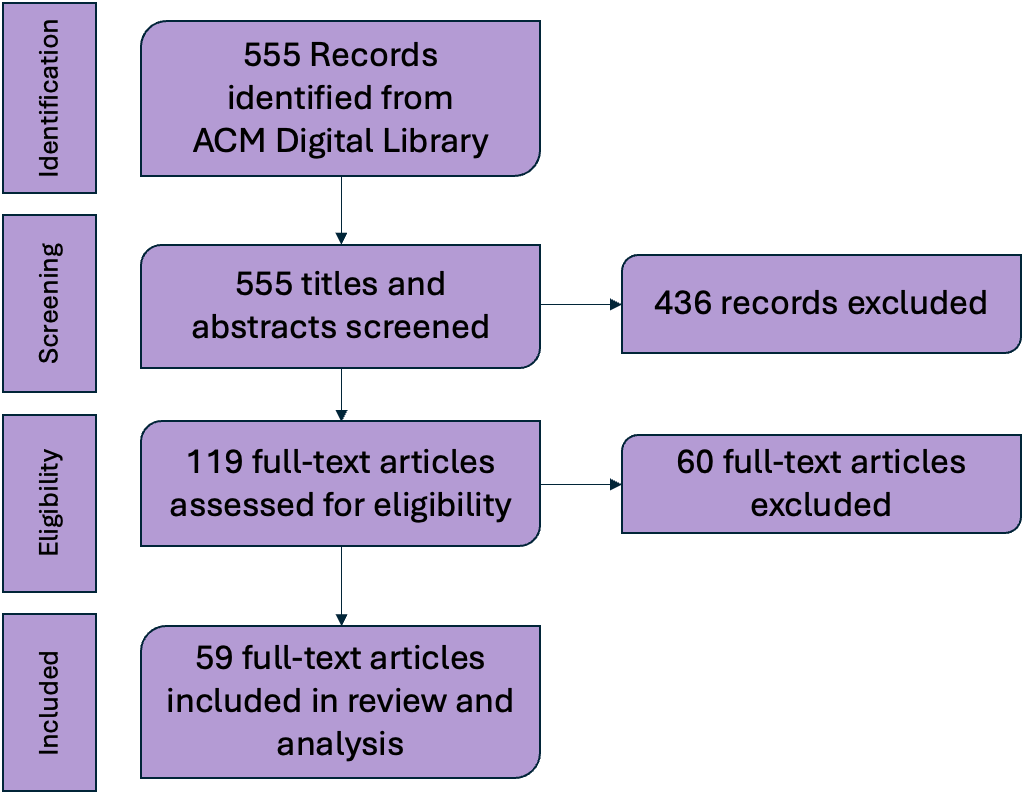}
    \caption{PRISMA flow diagram detailing the corpus of research articles screened and analyzed in this work. We started with 555 records from the ACM digital library (including 4 identified manually from the AIES '24 proceedings), removed 436 records during the abstract screening phase, and removed 60 full-text articles during the PDF eligibility assessment phase. This resulted in a final dataset of 59 full-text articles included in our review and analysis.}
    \label{fig:enter-label}
\end{figure}

Two authors conducted the abstract screening of the 555 articles identified through our search. This was conducted in multiple rounds of 50 papers each until a high inter-rater reliability score was reached for inclusion of works with a Cohen's Kappa of $0.94$. After this agreement was attained, one author coded the remaining 355 abstracts for inclusion.

Of the original 555 records, we excluded 436 papers during this stage. The exclusion criteria is detailed in Section \ref{sec:inclusion_criteria}. Of these exclusions, the vast majority were due to not being about scenario building ($n=183/436; 42\%$) but rather some type of analysis mentioning scenarios. Often times this captured ``scenario-based'' designs such as a predetermined set of variables or characteristics a self driving car might encounter for the purposes of testing; not futuring scenarios to consider impacts. The second most common reason to exclude was for works that were not full-text research articles ($n=167/436;38\%$); these typically took the form of extended abstracts or works in progress published in the ACM digital library, though some were proposals or descriptions of workshops or panels. %Sometimes the results of the workshops or sessions would have been relevant to this work, but the paper included in the library was only the proposal of the workshop. 
The third most common exclusion reason during the abstract phase was the subject matter not being about an emerging technology ($n=76/436;17\%$). The remaining three reasons for exclusion were scenario building being mentioned offhand but not the actual focus of the work ($n=6/436;1\%$), scoping reviews or SLRs ($n=2/436;0.5\%$), and finally fictional technologies that do not yet exist ($n=2/436;0.5\%$). %It is important to note we coded for exclusion on a hierarchical basis starting with formatively being a full research paper, then topically scenario building, then topically emerging technology. This means it was possible for a paper to be excluded for more reasons than one (e.g., an extended abstract that was not about scenario building and not concerning an emerging technology), but we only coded for the first exclusion criteria in our hierarchy; as a result these percentages of exclusion criteria are mutually exclusive and lower bounds. 

\subsection{Full Text Screening}

We next obtained the PDFs of the remaining 119 full-text articles and assessed them for eligibility, ultimately excluding 60 papers during this stage. Here we identified a new category for exclusion: speculative design works that aimed to iteratively improve on specific design features of existing technologies or prototypes through querying potential users rather than focusing on future impacts of the technology on communities or society at large. Though they sometimes used an adjacent technique to scenario building, the goal of these works was to result in technology adaptations more suitable to specific user bases, such as works that conducted interviews with their potential users to understand how they would like to see prototypes altered. For example, one work used scenario-based HCI methods to create design recommendations for cemetery design \cite{hirsch2022design} and another explored methods to make genAI code more explainable \cite{sun2022investigating}. We excluded $n=7/60$ articles for this reason. Our most common exclusion criteria was again works that were not about scenario building ($28/60;47\%$), followed closely by works not considering emerging technologies ($24/60;40\%$). We also excluded one paper ($1/60;2\%$) for not being written in English. %(it was written in Portuguese), which we missed during the abstract screening phase because the abstract was written in English. 
There were no other reasons for exclusion during this stage. 

\subsection{Descriptive Statistics}

The final corpus for inclusion in this work consisted of 59 papers from the ACM digital library (2 of which were identified in the extra inclusion of the AIES '24 proceedings). The conferences with at least 3 papers in our final database (no journals met this criteria) were:
\begin{enumerate}
    \item CHI: Human Factors in Computing Systems ($n=14/59; 24\%)$
    \item DIS: Designing Interactive Systems ($n=9/59;15\%)$
    \item CSCW: Computer-Supported Cooperative Work \& Social Computing ($n=6/59;10\%)$
    % \item AIES: Artificial Intelligence, Ethics, and Society ($n=2/59;3\%)$
    % \item ASSETS: Conference on Computers and Accessibility ($n=2/59;3\%)$
    % \item COMPASS: Conference on Computing and Sustainable Societies ($n=2/59;3\%)$
    % \item OzCHI: Australian Conference on Computer-Human Interaction ($n=2/59;3\%)$
    % \item Mindtrek: International Academic Mindtrek Conference ($n=2/59;3\%)$
\end{enumerate}

For the qualitative analysis, we iterated among three authors until we agreed on standard definitions for all variables for which we were coding. During this stage we coded for the type of emerging technologies studied, %the paper domain or topic focus, 
the dominant terminology used to describe scenario building, the scenario generation process (e.g., who wrote them, method to generate), the scenario evaluation process (e.g., how many evaluators, evaluator expertise), content of the scenarios, metadata of scenarios (e.g., medium, language), %content valence (focus on harms or opportunities), 
their goal of using scenario building, and degree of participatory design. 

\begin{figure}
    \centering
    \includegraphics[width=1\linewidth]{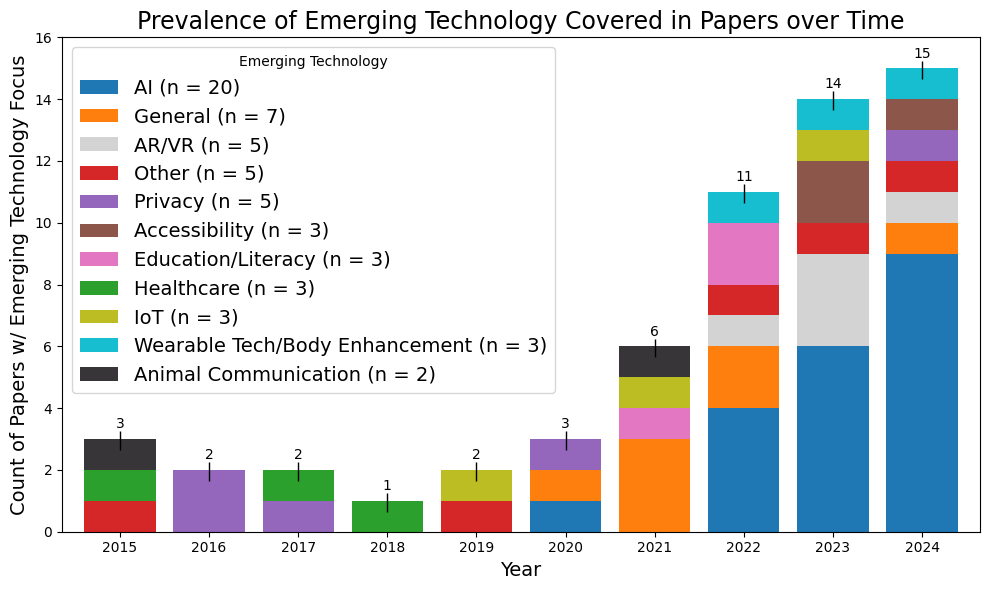}
    \caption{Stacked bar plot displaying the count of papers with a focus on the categorized emerging technologies over time. Works were categorized into ``other'' if only one paper covered that topic.}
    \label{fig:et-barplot}
\end{figure}

As shown in Figure \ref{fig:et-barplot}, there has been a recent and growing surge of the use of scenario based methods in computer science for works concerning emerging technologies. 11, 14, and 15 papers used these methods in 2022, 2023, and 2024, respectively, compared to an average of 2 papers per year from 2015-2020. We categorized the type of emerging technology that was the focus of these works, and grouped them in ``other'' if only one work discussed the topic. Of the 59 papers in our final corpus, $21/59; (36\%)$ were predominantly about AI, with all but one of those papers published in 2022 or later. Aside from ``general'' emerging technologies and ``other,'' our next most common emerging technologies were augmented reality/virtual reality (AR/VR) and privacy focused technologies, which tied at $5/59; (8\%)$ papers each discussing these topics. All of the AR/VR papers were published in 2022 or later, whereas all of the privacy focused papers were published in 2020 or earlier with $80\%$ of them published between 2016-2017. 

The terminology that researchers ascribe to this method is diverse, with six different nomenclatures present in our corpus. ``Speculative design'' was the most popular terminology, with $33/59; (56\%)$ papers predominantly referring to this method by that term, followed by ``scenario-based'' ($n=19/59;32\%$). The other four terms dominantly used for the method in our corpus each appeared in 2 papers: ``scenario building,'' ``scenario writing,'' ``speculative fiction,'' and ``what-if scenarios.''  There were no notable temporal trends, other than ``scenario writing'' and ``scenario building'' did not appear in computing literature until recently (once in 2021 and three times in 2024). Otherwise there has been a consistent usage of both ``speculative design'' and ``scenario-based.'' It is important to note that speculative design is used as a much larger umbrella term that covers non scenario-based methods. ``Scenario-based'' similarly includes a large body of work that deals with more variable setting simulation based works, such as those defining specific scenarios for testing in autonomous vehicles or mathematical models. To distinguish this distinct narrative style use of scenarios for futuring analyzed in this work from the larger existing bodies of speculative design and scenario-based methods, we decide to move forward with the use of ``\textit{scenario building}'' to describe this method, though note that all of these terminologies could mean the same thing in various settings.

% Layers for data exploration we can also add here:
% \begin{itemize}
%     \item Language
%     \item Medium
%     \item How many scenarios
%     \item How many people writing
%     \item Valence (focus on harms/opportunities)
% \end{itemize}

\section{Analysis}

The goal of this paper is to examine how researchers have utilized scenarios in computing research to meet various needs in order to provide a comprehensive snapshot of these types of methods in the field. In this section we first go through the objectives of using scenarios in the works in our corpus, as well as the form their scenarios took in terms of medium, content, and format. We then take a deeper look at the participatory element enabled by using scenarios by examining the generation process, the evaluation process, and the level of participation from lay stakeholders.

\subsection{Thematic Analysis: How are scenarios being used in computing literature?}

% \begin{itemize}
%     \item RQ1 objective column
%     \item cascaded back across the topic columns, generation process columns, terminology etc
%     \item use the other variables we have coded for in a thematic filtering analysis
% \end{itemize}

We seek to understand the objectives that researchers seek to achieve through the use of scenarios in computing literature. Of the 59 papers included in our final corpus, we qualitatively analyzed the objectives of using these scenarios as stated explicitly by the researchers (either in the form of research questions, contributions, or in their methods and analysis sections). We identified through inductive analysis five main categories to group these objectives; %, and assigned each paper a primary objective and most a secondary objective. 
%We define those categories 
in order of frequency they are: (1) \textbf{gather stakeholder needs and values} ($n=22/59; 37\%)$; (2) \textbf{empower marginalized groups to imagine technology futures} ($n=12/59; 20\%$; (3) \textbf{provoke ethical reflection and promote critical awareness} ($n=10/59; 17\%$); (4) \textbf{anticipate threats and risks of these technologies} ($n=10/59;17\%$); and (5) \textbf{explore perceptions and impact of novel technologies} ($n=5/59; 8\%$). %We now explore these more in-depth individually. 

\subsubsection{Gather stakeholder needs and values (Table A\ref{tab:obj-1})}

We define this objective for works that primarily used scenario building to understand the desires and needs of stakeholders so that they can inform design of technology or policy that would govern the use of these technologies in society. Scenario building is an effective method for this goal since researchers are able to contextualize complex technologies and concepts in short narratives to frame stakeholders with all the needed information to make an informed opinion in a short period of time. This was by far the largest category of use in our corpus with $n=22/59$  papers. %, and it was coded as the secondary objective for 8 additional works. 
These works were, especially in recent years, typically using scenario building to gather stakeholder needs and values around various AI technologies ($n=12/22)$, though people have used this method for many other emerging technologies as well.  

The terminology used for this objective was predominantly ``scenario-based'' ($n=9/22$), though ``speculative design'' closely following ($n=8/22)$, and every other terminology was used except for ``speculative fiction.'' %In terms of scenario medium, people using scenarios for this objective largely followed the overall trend of the corpus. The vast majority of the works used text as at least one medium for the scenarios ($n=19/22;86\%$), and of those 19 more than half ($n=10$) had text as the only medium. More than half the works ($n=13/22;59\%$) only used one medium, using text ($n=10$), video ($n=2)$, and speech ($n=1$), though when using multiple media, people additionally used images ($n=5$) and physical artifacts ($n=3$). 
In line with being the largest by frequency of use  objective, this was also the most diverse in terms of specific usage types. Some papers used is a springboard to gather stakeholder values about technology ethics, such as understanding Japanese citizens' perceptions of the Metaverse \cite{hohendanner2024metaverse} or to map public perception of AI chatbots \cite{kieslich2024my}. As \citet{hohendanner2024metaverse} put it: this method ``can assist with placing the needs and doubts of the actual user of the technology at the center of design decisions.'' 

Researchers also pursued this objective in the context of healthcare, well-being, and care-focused technology futures. In these works, they focused on a targeted group of stakeholders to evaluate the scenarios, such as understanding older adults' expectations of how a voice assistant should behave when providing health information \cite{brewer2022empirical}, or the use of chatbots to meet the healthcare needs of migrant workers \cite{tseng2023understanding}.

% Sometimes people in this objective used scenarios specifically to understand how certain technologies would be perceived or adopted in various types of education contexts. One study used scenarios to assess perceived usefulness, privacy concerns, and likelihood of adoption of learning analytics among instructors \cite{li2022scenario}. Another study sought to understand the potential benefits and harms of various technological interventions for education in Pakistan, and in their study they consulted students in tandem with workers at various education NGOs or other international development bodies \cite{khan2021speculative}.  

Some of these works consulted specific end users (rather than the general public) to understand how they perceived the risks and benefits of existing technologies, such as one method that used a Dungeons and Dragons style scenario method with journalists to understand their perception of possible deepfake identification technology \cite{sohrawardi2024dungeons}. Another consulted social housing tenants to understand their particular security concerns in relation to smart home devices %and how they imagined the future of living with them 
\cite{benton2023location}. In that vein, another study used scenarios to gain a deeper understanding of the perceptions of Australians with green energy technologies in their homes towards data sharing of everyone in their community in regards to privacy concerns and public responsibility for curtailment \cite{snow2021neighbourhood}.

The uniting thread among these works is that they were interested in the values of non technology stakeholders---whether that was end users, lay stakeholders, or key players such as journalists. They utilized scenarios to frame these technologies in a way that was easily understood by these stakeholders so that they could make informed and pointed value judgments on the perceived future implications of these technologies, whether positive or negative. 

\subsubsection{Empower marginalized groups to imagine technology futures (Table A\ref{tab:obj-2})}

We define this objective for works that used scenarios specifically to empower a marginalized group through speculative and participatory methods to understand the specific concerns, needs, and visions of these communities. These projects worked specifically with and/or for communities that are typically left out of larger discussions surrounding technology futures. The most common were works that dealt specifically with BIPOC communities ($n=5/12$), followed by communities of people with chronic diseases or disabilities ($n=4/12$), followed by works dealing with indigenous communities ($n=2/12$), and finally one work that focused on Iraqi refugees in Australia ($n=1/12$).%, such as indigenous groups, BIPOC communities, or people with disabilities. This was the second largest objective in our corpus, with $n=12/59  (20\%)$ papers falling into this category. 

Works in this category predominantly referred to the method as ``speculative design'' ($n=11/12$), with one paper referring to it as a ``scenario-based'' method. %Works using scenario building to empower communities had a higher usage of image as a medium, with $5/12;(42\%)$ works using images in addition to another medium (typically text) for their scenarios. They also had a slightly higher tendency to use more than one medium. %Compared to the larger corpus, works using scenario building for this objective had a much higher proportion of non-expert-written scenarios ($36\%$ vs. $17\%$). 
These works often focused on ``general'' emerging technologies and technology futures ($n=5/12$), but some also dealt with accessibility focused new technologies ($n=3/12$), and then one each focused on AI, green technology, education empowering technologies, and healthcare technology. Instead of focusing on different technologies, these works tended to focus on different communities. %Different from the first objective (gathering stakeholder inputs), the works using scenario building to empower marginalized communities all had a similar usage type and style deployed in their methods---the main differences were in the communities they focused on. The most common were works that dealt specifically with BIPOC communities ($n=5/12;42\%$), followed by communities of people with chronic diseases or disabilities ($n=4/12;33\%$), followed by works dealing with indigenous communities ($n=2/12;17\%$), and finally one work that focused on Iraqi refugees in Australia ($n=1/12;8\%$).

Within the works for BIPOC communities, these papers often referred to the concept of ``Afrofuturism,'' which artist Sanford Biggers describes as ``a way of re-contextualizing and assessing history and imagining the future of the peoples of the African Diaspora via science, science fiction, technology, sound, architecture, the visual and culinary arts, and other more nimble and interpretive modes of research and understanding'' \cite{winchester2018afrofuturism}. \citet{harrington2021eliciting} point out that ``Marginalized populations are very rarely, if ever, represented in
popular scenarios of technology design fictions ... the needs of white affluent citizens from financially-wealthy countries are at the center of such scenarios.'' \citet{bray2021speculative} describe using this method to empower Black and brown communities as ``an accessible resource that can be readily used by designers and nondesigners alike in addressing community concerns.'' Beyond focusing on BIPOC communities, these works also are likely to highlight other less dominant groups such as Black women, Femmes, or Non-Binary people \cite{klassen2024black}, or working-class Detroiters \cite{lu2024contamination}.

Within the works focusing on using scenarios to empower communities with chronic illness or disabilities ($n=4/12$), one focused generally on people with chronic illnesses \cite{hoang2018can}, one focused generally on people with disabilities and their access to technology \cite{hsueh2023cripping}, one focused on blind and low vision users \cite{phutane2023speaking}, and one focused on autistic Twitch livestreamers \cite{mok2024building}. All four of these used at least one non-text medium, with the work for blind and low vision users exclusively using speech scenarios. As \citet{hsueh2023cripping} put it, ``Disabled people have long been doing critical design – they critique existing built environments and (re)make them in ways that are often rendered invisible in history;'' they use scenario building in their work to make those designs visible. This type of work demonstrates the versatility of scenarios and how they can empower different people to take part in designing their own desired technological futures. 

There were two works that focused on indigenous communities in this section: one worked with an indigenous group in Northwestern Namibia and Southwestern Angola called ovaHimba \cite{muashekele2023ancestral}, and the other worked with Imazighen, the native people in remote Morocco \cite{ruller2022speculative}. Both works took extreme care to fully situate themselves within the context of these communities, and both projects are working with established collaborations between the researchers and the indigenous communities that have been going on for many years. The work with ovaHimba sought to understand the citizen desires of the use of green energy through group ``future walk'' studies where they would collectively take a walk through their communities and discuss desired potential technology to help them \cite{muashekele2023ancestral}. The other focused on illiteracy and ways to ameliorate the negative effects of this through new technologies \cite{ruller2022speculative}. 

% Do I need to do a short paragraph about the one paper that focused on Iraqi refugees? \cite{almohamed2020magic} (Kimon let me know what you think.)

\subsubsection{Provoke ethical reflection and promote critical awareness (Table A\ref{tab:obj-3})}

We define the objective for this group of works to mean the use of scenarios to prompt people to think through social and ethical ramifications of technologies, question norms, and imagine other possible futures. This was tied for the third most common objective, with $n=10/59 (17\%)$ works having this as the primary objective for their use of scenarios. 

Works in this group also favored the terminology ``speculative design'' ($n=6/10$), but it was also referred to using ``scenario-based'' ($n=2/10$), ``scenario building'' ($n=1/10$), and ``speculative fiction'' ($n=1/10$). %Almost all of them ($90\%$) used text as at least one of the mediums, though of the multi-media scenarios, $n=3/5; 60\%$ used images and $n=2/5; 40\%$ used audio or speech. 
There were three works in this category focusing on wearable technology/body enhancement technology ($n=3/10$), two each that focused on AI, animal communication technologies, and privacy focused technologies ($n=2/10$ each), and one that focused on emerging technologies in general. %These works all tended to focus on the technology itself rather than the specific implications in various domains such as community focused or healthcare focused. 
Works in this category focused on different methods to promote reflection of technology ethics and utilized the method as a way to surface ethical issues of technologies early.

One work used the popular television show \textit{Black Mirror}, a science fiction dystopian show that focuses on dark potential impacts of technology, to cultivate ethical reflection in computer science eduction \cite{klassen2022run}. In a similar vein, another work used the concept of ``cinematic pre-visualization (previs)'' techniques---they assert ``using medium-fidelity previs methods
can result in compelling visual artifacts that propel researchers’ vision for their work, highlight concrete challenges and opportunities, and facilitate public discussion'' \cite{ivanov2022one}. Another work analyzed how researchers could make use of ``Model Cards'' (documents providing benchmark evaluation information about trained machine learning models) to promote ethical reflection \cite{nunes2022using}. They found that while these cards prompted participants to ``deliberat[e] on the ethics of their development decisions, they only recorded those that they considered to be ethical.'' 

Some works in this category focused on %the overall human perception of human augmentation%---a term they use to describe ``technologies that help users overcome their limitations and integrate with the human body'' \cite{villa2023understanding}. %They used different variants of scenarios, those focusing on different types of augmentation, to 
understanding the driving factors behind different perceptions of %these 
human augmentation technologies \cite{villa2023understanding} or %Another work focused on 
``personal fabrication technologies,'' %which are not exclusively though commonly human augmentation artifacts, 
and the latter even used a narrative scenario in their research paper to educate the reader on how they define this technology and contextualize target users \cite{stemasov2022ephemeral}. % Another work focused on emerging wearable technologies that seek to mediate intimacy over long distances, and constructed fictional stories to explore the ``potential benefits, effects, challenges, and shortcomings'' of these technologies \cite{aljuneidi2023disclose}.
%Finally, two 
Other works focused on the set of technologies aiming to facilitate communication with pets and provoke reflection on this %technologies; %. Both were slightly older papers, with one published in 2015 and the other in 2021. %The first compared experts in animal welfare with (comparably ignorant) pet owners, and found a stark mistmatch between the values and concerns of experts in animal welfare and non-expert pet owners concerning the technology.
%``a strong desire among pet owners for technology that has little scientific justification, whilst our experts caution that the use of technology to augment human-animal communication has the potential to disimprove animal welfare'' \cite{lawson2015problematising}.
%They asserted their goal ``was to 
%``explore 
``hypothesized future technology; our intention was to raise awareness of issues with products before they become public concerns, not
afterwards'' \cite{lawson2015problematising}, and found %. The second work sought to provoke reflection 6 years later as technology for pets became more commonplace, and found that  `
``speculative design proved to be a useful tool for provoking ideas and framing discussions around this theme'' \cite{french2021ethics}.

\subsubsection{Anticipate threats and risks of these technologies (Table A\ref{tab:obj-4})}

We classified works into this objective when they primarily used scenarios as a method to anticipate potential threats of technology, security exploits, malicious uses, or other harms that could arise from the technology. This was tied for the third most common main objective, with $n=10/59;(17\%)$ works having this as the primary goal.

Works with this goal predominantly referred to the method as ``scenario-based'' ($n=5/10$), with ``speculative design'' falling in close second ($n=4/10$), and one work using the term ``speculative fiction'' ($n=1/10$).  %Almost all of these works utilized scenarios written either entirely by experts ($n=7/10$;70\%) or by a combination of experts and non-experts ($n=2/10;20\%$), with only one using scenarios solely written by non-experts ($n=1/10;10\%$). Every single work in this category utilized some form of text scenario, with some also having images ($n=2/10;20\%$), audio ($n=1/10;10\%$), and one having virtual reality scenarios ($n=1/10;10\%$).
These works followed a trend of anticipating ``hot topic'' technologies of their year. One paper in 2019 focused on anticipating threats of internet of things (IoT)/smarthome technologies, followed by one in 2020 focusing on privacy. From 2022-23 there were three papers focusing on augmented reality/virtual reality (AR/VR) technologies, and then in 2024 there were four papers focusing on the threats of AI. These works all tried to anticipate threats of specific technologies before all of their harms have been realized in society.

One work focused on the specific risks of children utilizing IoT technology in order to anticipate the privacy, security and safety risks that emerge in the context of such uses \cite{knowles2019scenario}. %They let children play around with the technology, and then based on those experiences wrote up scenarios narrativizing some of the risks they discovered through this process in order to present the risks to a wider group. 
Another work wove scenarios into a role-playing game to help software developers identify security threats \cite{merrill2020security}. As the author asserts, ``threat identification is
a socially-situated practice. Multiple stakeholders and values collide with business imperatives to produce a socially contingent set of threats deemed relevant,'' which scenario based methods are perfect at eliciting. %One other work in this category used scenarios to identify ``hazards'' and risks of socio-techical systems in HCI work \cite{watson2021hci}.

The works on AR/VR all shared a concern that risks were unrecognized in these technologies in tandem with them becoming more pervasive. One used ``dark scenarios'' to explore how it %``Ubiquitous Augmented Reality 
``may be used to create deceptive designs taking advantage of users, or could lead to unintentional, but still equally harmful negative consequences'' \cite{eghtebas2023co}. %Another used scenarios to analyze a much more focused harm of the potential for memory manipulation in AR/VR technologies \cite{bonnail2023memory}. 
Another created scenarios in VR to explore the potential malicious use of perceptual manipulation in VR in order to explore threats within this space \cite{tseng2022dark}.

The most recent of theses works focused on AI technologies and were all published in 2024. One used scenarios to generally understand AI researchers' concerns in regard to AI \cite{jantunen2024researchers}. %One explored how AI systems, especially LLMs that can be applied to many tasks, behave compared to humans in high-stakes military decision-making scenarios \cite{lamparth2024human}.
To illustrate how important scenarios are in making complex threats manifest in a much more concrete and digestible way, one utilized scenarios to help users foresee AI harms, and one of their participants remarked ``Why the heck didn’t we think like this before? I’ve used tons of AI Ethics tools. My clients hate how abstract they are. This [process] is specific and customers would love it...this is the next leap from impact or risk assessments'' \cite{ehsan2024seamful}.

\subsubsection{Explore perceptions and impact of %and sometimes prototype 
novel technologies (Table A\ref{tab:obj-5})}

The final objective was defined as works that used some sort of artifact to explore people's perceptions of novel technologies either through simulation of digital and physical artifacts or detailed descriptions of what they might look like. There were only $n=5/59 (8\%)$ works that fit this category, and they all focused on different technologies. Most of these works used the terminology ``speculative design'' ($n=4/5$), and one used the term ``scenario-based'' ($n=1/5$). Two of these works used videos as the primary form of scenario in their works ($n=2/5;20\%$), which is proportionally twice as often as the larger corpus.

%All of these works explored an entirely new technology and provided people with some sort of demonstration of the technology, whether as a physical mock up or digital interface. 
One explored what intimate artifacts could look like in the future \cite{kaur2022future} by encouraging users to design scenarios detailing them. Another used an ``ethnographically-informed Walking \& Talking method'' to brainstorm possibilities of a new technology, in which they prototyped an ``interactive smell-stick device'' to share smells as you would photos on a smart phone \cite{stals2019urbanixd}. Others explored a microchip-based contraceptive implant \cite{homewood2017turned}, an anti-harassment system for social networks \cite{kim2024respect}, and finally one explored both utopian and dystopian body perception transformation technologies \cite{turmo2023futuring}.

Researchers discussed how using scenario methods for these novel technologies ``expanded the horizons...from what we had initially conceived from our secondary research'' \cite{kaur2022future}. Others noted that this method ``revealed key elements and design goals to consider'' \cite{kim2024respect}. Another work mentioned that ``although described in the popular press and articles in the medical field, the technology needs to be critically examined from the perspective of how we will interact with it, and the role it may play in our lives'' \cite{homewood2017turned}. These all suggest that the use of scenarios in this manner can elucidate findings that other traditional methods cannot.

\subsection{The Participatory Element of Using Scenarios}

A recent work by \citet{delgado_participatory_2023} examining AI design detailed the levels of participation usually applied in ``participatory designs,'' ranging from (1) consult, (2) include, (3) collaborate, and finally (4) own. We use these levels to analyze the manner in which scenario building in our corpus was used to achieve participatory means, especially in tandem with the level of expertise of those writing the scenarios. We assess the participation in relation to how stakeholders are involved in the general process of the scenario method---depending on the structure of the task, they were typically involved in creating, evaluating, or reflecting upon the scenarios. We provide our own detailed definitions used to apply these levels below:

\begin{enumerate}
    \item \textbf{Consult:} ranking, feedback, or multiple choice type evaluation, such as those that had people deliberate about pre-defined policy options.
    \item \textbf{Include:} the options are originally created by the researchers, but participants have the ability to shape the options and alter them or provide meaningful feedback that can change outcomes. Though they are not part of the creation process, they still have the power to shape the emerging technologies.
    \item \textbf{Collaborate:} participants are part of the creation process, but they still do not have equal authority or power relative to the researchers.
    \item \textbf{Own:} participants are central to the process, co-design everything, and have outcome authority. There is an equal partnership between researchers and participants, if not skewed more towards the participants. 
\end{enumerate}

\begin{figure}
    \centering
    \includegraphics[width=1\linewidth]{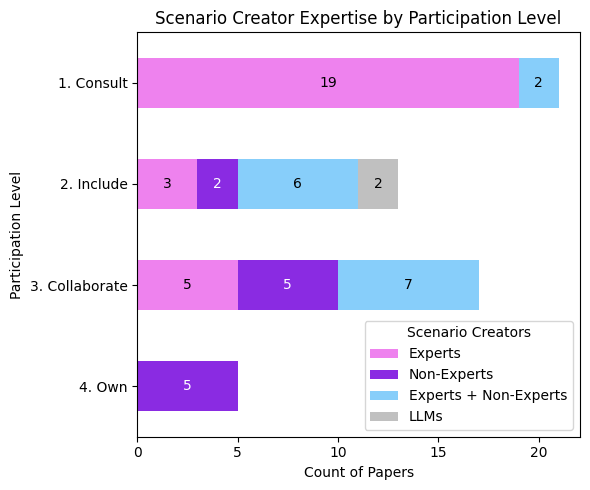}
    \caption{Stacked bar plot of the 56 papers in our corpus with participatory elements categorized by participation level and separated by expertise of the scenario creators.}
    \label{fig:participation-plot}
\end{figure}

In this section we also delineate between the inclusion of expert and non-expert stakeholders. We define experts to be users or participants with a professional background (either technical or domain based) in the technology that is being evaluated. Non-experts are people who have no formal training in either case, though they possess their own contextualized lived experience which constitutes its own form of situated expertise. An overview of the categorization of works in this section can be seen in Figure \ref{fig:participation-plot}. There were only three papers in our corpus without a participatory element (the scenarios were written and evaluated by the authors alone), %: \citealt{hsueh2023cripping, stemasov2022ephemeral, digmayer2015designing}), 
leaving us with $56$ to analyze in this section.

\subsubsection{Level 1: Consult}

%This category includes works that \textit{consults} stakeholder involvement through ranking, feedback, or multiple choice type evaluation, such as those that had people deliberate about pre-defined policy options. 
This was the most common level of participation, with $n=21/59$ works in our corpus meeting this level. These scenarios were almost entirely written by experts ($n=19/21$), with $n=2/21$ being written by a combination of experts and non-experts. However, in instances in which the creators of the scenarios were not the evaluators of the scenarios, ($n=14/21$), non-experts tended to be the evaluators of the scenarios ($n=9/14$). 

These works took many forms, from eliciting ethical concerns from AI researchers and exploring how users reason about future consequences \cite{jantunen2024researchers}, to exploring how specific technology could aid or hinder coping with interpersonal racism \cite{to2022interactive}, or to understand older adults' expectations of how a voice assistant should behave when providing health information \cite{brewer2022empirical}. The goal of this form of participation was to understand users' and stakeholders' perceptions of the impact of these technologies---not to iteratively create a better form of a specific design, but to understand in a larger sense how these technologies are viewed and impact lay stakeholders.

\subsubsection{Level 2: Include}

%The next level of participation is that which \textit{includes} stakeholder input. In this case the options are originally created by the researchers, but participants have the ability to shape the options and alter them or provide meaningful feedback that can change outcomes. Though they are not part of the creation process, they still have the power to shape the emerging technologies. 
This was the third most common category for works in our corpus, with $n=13/59$ works meeting this criteria. These scenarios had a wide disparity of creators, with $n=3/13$ written by experts, $n=2/13$ written by non-experts, $n=6/13$ written by a combination of the two, and $n=2/13$ written by large language models (LLMs). Notably these were the only two papers in the entire corpus that used LLM created scenarios, and both did so for participatory means.

Some examples of works in this category include exploring and illustrating possible AI extended reality threats and harms to guide future mitigation efforts \cite{baldry2024embodied} and uncovering children and caregivers' concerns for child-targeted navigation apps \cite{silva2022addressing}. \citet{khan2021speculative} describe using this method ``to understand the hurdles and possibilities for technological interventions that could help curtail the challenges facing the education sector and critically reflect on design possibilities and their anticipated consequences in order to
work towards positive outcomes in the future.'' Works meeting this inclusion criteria of participation afforded more agency to the participants---as \citet{benton2023location} note, their ``participatory approach to speculative design...allow[ed] opportunities for tenants to adapt
the initial technological future proposed by the research team and
then subsequently speculate about how this imagined future might
manifest within their own domestic lives.'' Though the creation process was owned by the research team, the participants still had power to shape their technological futures. 

\subsubsection{Level 3: Collaborate}

%In the \textit{collaboration} category, participants are part of the creation process, but they still do not have equal authority or power relative to the researchers. 
This was the second most common level of participation in our corpus, with $n=17/59$ works meeting this criteria. These works had a perfectly equal distribution of expert/non-expert scenario creators, with $n=5/17$ each being written by experts and non-experts, and the remaining $n=7/17$ being written by a combination of the two. The majority of these works had the creators of the scenarios also serve as the evaluators of the scenarios ($n=12/17$), and most of those that did not had non-experts evaluate the scenarios ($n=4/5$). 

This category had the highest proportion of works (relative to the greater corpus) in which the authors wrote scenarios based on the observed and discussed experiences of the stakeholders for which they were conducting research ($7/17;41\%$). It also had the highest proportion of works that used a workshop method to generate the scenarios ($n=12/17;71\%$, compared to the corpus of $20/59;34\%$). This for instance could take the form of watching how participants interacted with digital or physical artifacts and then writing up scenarios based on those \cite{hohendanner2023exploring}, or the reverse, where participants brainstormed out loud as a group and then the authors created digital artifacts and scenarios to represent those \cite{klassen2024black}. 

Some examples of works enabling this level of participation include one that empowered ``marginalized communities---especially Black and LatinX youth---to envision and co-create equitable, identity-affirming futures through speculative design grounded in Afrofuturism'' \cite{bray2021speculative}. Another described the method as useful to ``help practitioners identify AI system weaknesses to enhance explainability \& support user agency'' \cite{ehsan2024seamful}. One other described the goal as to ``gain a deeper understanding of deepfake detection tools and to offer guidance for the development of a reliable deepfake detection tool that empowers journalists to validate media sources, thereby enhancing the quality of their reporting'' \cite{sohrawardi2024dungeons}.

\subsubsection{Level 4: Own}

%In the \textit{ownership} category, participants are central to the process, co-design everything, and have outcome authority. There is an equal partnership between researchers and participants, if not skewed more towards the participants. 
This was the rarest level of participation in our corpus, with only $n=5/59$ works meeting this criteria. Every work that fit the \textit{own} criteria had solely non-experts write or create the scenarios. One of these worked with two indigenous communities in northwest Namibia and Southwestern Angola, and empowered this group to brainstorm their own technology futures with some facilitation from the researchers \cite{muashekele2023ancestral}. Another worked with young Black adults to co-design futures they imagined \cite{harrington2021eliciting}. By enabling participants to \textit{own} the process, \citet{lu2024contamination} discuss using participatory speculative design as a ``participatory space to disrupt the large-scale sociotechnical imaginaries shaped by institutions and markets and to allow for the emergence and articulation of community-held sociotechnical imaginaries that impacted communities can act upon.''

\section{Discussion}

% [prescriptive role] add a direction for where the field ought to go; write the research agenda for what needs to be done
% talk about the heuristic value of what we found and then describe where it should go based on this
% Look at what other fields are doing and draw analogs from them

These findings highlight that deploying scenarios in computing is a versatile approach that can effectively incorporate participatory elements. We found the most common reason to use scenario building was to gather stakeholder needs and values, though this also happened to coincide with the least involved form of participation: consulting stakeholders. Other popular objectives researchers sought through the use of this method, such as empowering marginalized communities to imagine their own technology futures, tended to coincide with higher levels of participation like collaboration or ownership of the process. Scenarios were also a popular method to provoke ethical reflection of emerging technologies or even anticipating specific risks of them.

20\% of the works in our corpus had exclusively non-experts create the scenarios, an additional 27\% had a combination of experts and non-experts create them, and another 17\% of the corpus has expert-written scenarios with non-expert evaluation. That makes 64\% of the entire corpus where non-experts played a crucial role in either creation or evaluation of the scenarios. Scenario building is a method particularly apt at integrating non-expert knowledge due to the ability to contextualize complex and abstract ideas into a narrative for them to then make informed opinions about potential futures, especially given the human propensity for storytelling \cite{gottschall2012storytelling}. These scenarios do not even have to be designed for a literate community---we saw a variety of mediums including non-textual storyboards designed for illiterate communities \cite{ruller2022speculative}, as well as audio scenarios designed for blind and low vision stakeholders \cite{phutane2023speaking}. 

An important element of the use of scenarios is determining \textit{who} writes, evaluates, or reflects upon scenarios. As \citet{carroll1995scenario} puts it, ``Our best course is to develop rich and flexible methods and concepts ... ideally by involving the users themselves in the design process.''  Thus, researchers highlight the relevance of lay-stakeholders, i.e. people who have no technical or domain expertise, but have a contextualized knowledge through lived experience that constitutes its own unique form of expertise. % that is essential to the conversation of risk assessment and technology futuring. %These insights can be informative for a multitude of purposes, for instance, to enhance risk assessment practices \cite{kieslich2025scenario}, develop technologies adapted to specific communities \cite{harrington2021eliciting}, or inform user-centered design of technologies \cite{benton2023location}. 
As means to bring lay-stakeholder input to life, an interesting trend among the works in this corpus was the author creation of scenarios through observations of the subject communities. $32\% (19/59)$ of the works in our corpus followed this practice. This follows a trend of scenario-based design noted in the 90s where researchers ``gather requirements by visiting their users' workplace, observing what was done and how it was done, even trying themselves to carry out the domain activities'' \cite{carroll1995scenario}, indicating that computing scholars continue to iterate upon old trends of the method.

Something the forebearers of scenarios may not have foreseen is the use of LLMs to write scenarios, yet two ($n=2/59$) works in our corpus did just that. Given their ability to produce comprehensive text, LLMs have the potential to write and re-write scenarios under any number of conditions and in high volume. As compact representations of actors, actions, events, and relationships, scenarios could almost be considered the ``file format'' of the social world, accessible to LLMs for manipulating in any number of ways. This allows for counterfactual exploration of a large space of possibility, while preserving legibility and salience which can facilitate stakeholder inclusion (e.g., via evaluation, surveys, stimulus, etc.) in different stages of the scenario building and evaluation process. In our corpus, one work used LLMs to create a large number of scenarios to iterate on the potential effectiveness of a policy condition under the EU AI Act \cite{barnett2024simulating}, and the authors later used these scenarios as a means to contextualize harms in order for lay-stakeholders to create informed suggestions for new policies \cite{barnett2025envisioning}. %The use of LLMs to create scenarios can isolate specific factors for which researchers wish to evoke responses, however this may come at the cost of the complexity and creativity of human created scenarios.

Our results show that the computer science literature adapted scenario methods close to the research tradition in the field of future studies \cite{borjeson_scenario_2006, van_der_heijden_scenarios_2005}. Thus, most papers analyzed in this SLR present multiple plausible futures of technology development \cite{amer_review_2013}. Two main goals of the use of scenarios in computing is to gather stakeholder feedback and include perspectives from marginalized communities, which is closely related to participatory foresight activities \cite{nikolova2014rise}. %The focus of studies in this SLR mostly lies on gathering stakeholder views with different means, for instance, in researching their needs, empowering marginalized groups, or explore risks and ethical reflections of emerging technologies. Thus, c
An aim of the use of scenarios in computing that diverges from traditional branches of the use of scenarios is to align technology development to the risk and benefit perceptions of stakeholders and imagine \& develop new technologies relying on stakeholder input. For instance, in business literature one of the central aims of scenario use is to engage in strategic planning that steers business decisions in desired, mostly profit-oriented, directions \cite{selin_trust_2006, varum_directions_2010}. In the anticipatory governance literature, scenarios are mostly used to test, develop and inform policy interventions \cite{brey_ethics_2017}.%, mittelstadt_how_2015, barnett2025envisioning}. 

The technology improvement centered use of scenarios might also explain the terminology used in many of the studies in this review. By and large, the emphasis lies on the term \textit{speculative design}, which mostly entails the imagination and design of new or adopted technologies. Consequently, scenario use in computing literature adopts the key methodological elements of scenario methods (e.g., development of plausible future pathways, inclusion of stakeholders), but adapted toward technology centered questions, e.g., how technology development can be altered to reflect the needs of the ones affected. There is potential to expand the application of scenarios in computer science by incorporating different branches of related scenario work. For example, the stated goals of conferences such as ACM FAccT and AIES are to study the societal, political, and regulatory contexts and impacts of computing technology. This could lead to more extensive use of scenario methods for regulatory and policy-related issues. While some studies using these methods have been published at these conferences \cite{ barnett2024simulating, barnett2025envisioning, kieslich2024my}, future work can build on this line of research to expand societal impact.

\subsection{Limitations}
While a central aspect and goal of this SLR was to evaluate a comprehensive body of work using scenarios, this work focused entirely on works published within the ACM digital library. As such, we inherently excluded venues outside of traditional computer science work that frequently use scenarios in the manner we describe, such as work in the social science literature \cite{diakopoulos_anticipating_2021, das_how_2024}, future studies \cite{borjeson_scenario_2006, van_der_heijden_scenarios_2005}, ethics/philosophy \cite{floridi_ethical_2020, schuijer_citizen_2021}, or legal studies \cite{helberger_four_2020, helberger_futurenewscorp_2024}. In a similar vein, we only included works focusing on emerging technologies to consider potential future impacts. As a result we excluded works that used this method for non-technological means or even mature technology that has reached a more stable state in society.% yet we can still anticipate impacts in new settings. 

We also chose to scope the time period for this work: with a 10 year horizon we might also miss connections back to earlier frames on the idea; for instance \textit{Scenario-Based Design}, a seminal work in this space, dates back to 1995 and contains many of the trends discussed in this work \cite{carroll1995scenario}, and other works in the early 2000s describe participatory design through storytelling \cite{muller_2002_participatory}, which further emphasizes the long tradition of this methodology in HCI research under different guises and with sometimes different nomenclature.  % We scoped our SLR on recent scenario usage in computing to specifically focus on novel technologies. 
%Adding to this, 
However, we found a surge in usage over the last three years (see Figure \ref{fig:et-barplot}) strengthening our observation that the use of scenario methods gained traction in the computing field in the last few years.

We also excluded 167 out of our initial pool of 555 articles for not being full research articles. A large portion of these were extended abstracts that would have been otherwise relevant to the analysis. Scenario building can be a useful early stage of risk assessment, and thus well suited for exploratory early works. Analyzing these works as well, though outside of our scope to evaluate full articles that have gone through the peer review process, may have revealed additional insights about using this method early in the risk assessment process.

\section{Conclusion}

In this work we conducted a systematic literature review of research papers in the ACM digital library over the last decade that worked with scenarios to consider the potential future impacts of emerging technologies ($n=59$). This work consolidated a fragmented space of scenario based methods within the computing literature and explored the various objectives for which researchers were using scenario building, as well as quantified the participatory element enabled by deploying the method. Future researchers can turn to this assessment of the scenarios in computing literature as a springboard for ways to deploy this method, as well as to understand the different objectives and participatory means that can be achieved through this type of work.

\section{Ethical Statement}

All authors declare no conflicts of interest for this work.

\subsection{Ethical Considerations Statement} 

We do not see a large potential for ethical issues in regard to this work because this was a review entirely of secondary data to detail an overview of the way a method is deployed. However, we made editorial and scoping decisions for which works to include in this review, and thus we present a biased interpretation of this body of work. Our work only includes those written in English, and though this is a reflection of the wider trend of English being the predominant language of published scientific works \cite{drubin2012english}, this certainly excluded any potential for non-English works such as the one we excluded in our review \cite{macedo2020vis2learning}. Further, by focusing on the ACM digital library alone, we perpetuate any biases encapsulated by this catalog of published work. 

Additionally, the coding was done qualitatively by two authors on this work. It is possible that others may disagree with the works we chose to include, and that could thus change the results we present. In order to  mitigate this potential concern, we provide all of our finalized coding documents at this link: \url{https://tinyurl.com/scenario-slr}.

\subsection{Researcher Positionality Statement}

This was an interdisciplinary collaboration, with two authors working at the intersection of computer science and communications, one working in communication and pursuing a career in the legal sector, and one identifying as a social scientist working as a postdoctoral researcher in an institute focusing on information law. Two of the authors identify as female, and two identify as male. Three of the authors identify as white, and one author identifies as Asian. As a predominantly non-BIPOC team, we took care to  analyze the section on marginalized communities such as BIPOC groups by reporting on how these communities use this method rather than any prescriptive analysis; we used direct quotes when appropriate rather than our own take on the subject. 

\subsection{Adverse Impact Statement}

We only captured methods utilizing scenarios in computing literature in this review. As we stated throughout our introduction, related work, and discussion, there are many other ways in which to use scenarios that were not covered in computing literature. One potential harm of this work is that people may see this as a prescriptive approach of the only ways to employ scenarios, thus stifling methodological creativity and excluding other established ways of using scenarios. We do not believe the methods and objectives covered in this work are the only ways that can or should be used, but we believe that is a potential negative impact of this work. Overall, we believe the benefits of this work far outweigh these potential risks.

% \section*{Acknowledgments}
% Acknowledgments have been blinded for review.

\bibliography{references}

\appendix
\setcounter{secnumdepth}{2}

\section{Appendix}

\begin{table*}[ht!]
    \centering
    \rowcolors{2}{LightSteelBlue1!50}{white}
    \begin{tabular}{clllll}
        \toprule
        \multicolumn{6}{c}{\textbf{Objective: Gather stakeholder needs, and values}} \\
        \toprule
        Year & Reference & Emerging Technology & Terminology & Scenario Creators & Participation\\
        \midrule
        2015 & \citeauthor{digmayer2015designing} & Other% : Urban Transportation 
        & Scenario-Based & Experts & N/A\\
        2015 & \citeauthor{hosseini2015discrete}  & Healthcare & What-If Scenarios & Experts & 1. Consult \\
        2017 & \citeauthor{sailaja2017challenges}  & Privacy & Scenario-Based & Experts & 1. Consult\\
        2020 & \citeauthor{kaur2020using}  & AI & Scenario-Based & Experts & 1. Consult\\
        2021 & \citeauthor{khan2021speculative}  & Education/Literacy & Speculative Design & Combination & 2. Include\\
        2021 & \citeauthor{snow2021neighbourhood}  & IoT & Speculative Design & Experts & 1. Consult\\
        2022 & \citeauthor{brewer2022empirical} & AI & Scenario-Based & Experts & 1. Consult\\
        2022 & \citeauthor{li2022scenario} & Education/Literacy & Scenario-Based & Experts & 2. Include\\
        2022 & \citeauthor{park2022social} & AI & What-If Scenarios & LLMs & 2. Include\\
        2022 & \citeauthor{silva2022addressing} & Other%: Child Locative Navigation
        & Scenario-Based & Combination & 2. Include\\
        2023 & \citeauthor{alfrink2023contestable} &	AI & Speculative Design & Experts & 1. Consult\\
        2023 & \citeauthor{benton2023location} &	IoT & Speculative Design & Combination & 2. Include\\
        2023 & \citeauthor{das2023algorithmic} &	AI & Scenario-Based & Experts & 2. Include\\
        2023 & \citeauthor{hohendanner2023exploring} &	AI & Speculative Design & Experts & 3. Collaborate\\
        2023 & \citeauthor{ringfort2023design} &	AI & Speculative Design & Combination & 3. Collaborate\\
        2023 & \citeauthor{tseng2023understanding} &	AI & Scenario-Based & Combination & 2. Include\\
        2024 & \citeauthor{barnett2024simulating} & AI & Scenario Writing & LLMs & 2. Include\\
        2024 & \citeauthor{bertrand2024ai} & AI & Scenario-Based & Experts & 1. Consult\\
        2024 & \citeauthor{hohendanner2024metaverse} & AR/VR & Speculative Design & Non-Experts & 3. Collaborate\\
        2024 & \citeauthor{kieslich2024my} & AI & Scenario Writing & Non-Experts & 3. Collaborate\\
        2024 & \citeauthor{kolovson2024using} & Privacy & Speculative Design & Combination & 2. Include\\
        2024 & \citeauthor{sohrawardi2024dungeons} & AI & Scenario Building & Experts & 3. Collaborate\\
        \bottomrule
\end{tabular}
\caption{\label{tab:obj-1}All works in our corpus with the primary objective of using scenarios to gather stakeholder needs and values. Organized chronologically by year, then alphabetically by reference.}
\end{table*}

\begin{table*}[ht!]
    \centering
    \rowcolors{2}{LightSteelBlue1!50}{white}
    \begin{tabular}{clllll}
        \toprule
        \multicolumn{6}{c}{\textbf{Objective: Empower marginalized groups to imagine technology futures}} \\
        \toprule
        Year & Reference & Emerging Technology & Terminology & Scenario Creators & Participation\\
        \midrule
        2018 & \citeauthor{hoang2018can} &	Healthcare & Speculative Design & Combination & 3. Collaborate\\
        2020 & \citeauthor{almohamed2020magic} &	General & Speculative Design & Non-Experts & 3. Collaboration\\
        2021 & \citeauthor{bray2021speculative} &	General& Speculative Design & Non-Experts & 3. Collaboration\\
        2021 & \citeauthor{harrington2021eliciting} &	General& Speculative Design & Non-Experts & 4. Own\\
        2022 & \citeauthor{ruller2022speculative} &	Education/Literacy & Speculative Design & Experts & 3. Collaborate\\
        2022 & \citeauthor{to2022interactive} &	General& Speculative Design & Experts & 1. Consult\\
        2023 & \citeauthor{hsueh2023cripping} &	Accessibility& Speculative Design & Combination & N/A\\        
        2023 & \citeauthor{muashekele2023ancestral} &	Other%: Green Technology
        & Speculative Design & Non-Experts & 4. Own\\
        2023 & \citeauthor{phutane2023speaking} &	Accessibility& Speculative Design & Experts & 1. Consult\\
        2024 & \citeauthor{klassen2024black} &	General& Speculative Design & Non-Experts & 3. Collaborate\\
        2024 & \citeauthor{lu2024contamination} &	AI& Speculative Design & Non-Experts & 4. Own\\
        2024 & \citeauthor{mok2024building} &	Accessibility& Scenario-Based & Experts & 1. Consult\\
        \bottomrule
\end{tabular}
\caption{\label{tab:obj-2}All works in our corpus with the primary objective of using scenarios to empower marginalized groups to imagine technology futures. Organized chronologically by year, then alphabetically by reference.}
\end{table*}

\begin{table*}[ht!]
    \centering
    \addtolength{\tabcolsep}{-0.2em}
    \rowcolors{2}{LightSteelBlue1!50}{white}
    \begin{tabular}{clllll}
        \toprule
        \multicolumn{6}{c}{\textbf{Objective: Provoke ethical reflection and promote critical awareness}} \\
        \toprule
        Year & Reference & Emerging Technology & Terminology & Scenario Creators & Participation\\
        \midrule
        2015 & \citeauthor{lawson2015problematising} & Animal Communic. & Speculative Design & Experts & 1. Consult\\
        2016 & \citeauthor{van2016computationally} &	Privacy & Speculative Design & Experts & 1. Consult\\
        2016 & \citeauthor{zhang2016opportunities} &	Privacy & Scenario-Based & Experts & 1. Consult\\
        2021 & \citeauthor{french2021ethics} &	Animal Communic. & Scenario Building & Combination & 3. Collaborate\\
        2022 & \citeauthor{ivanov2022one} &	General & Speculative Design & Combination & 3. Collaborate\\
        2022 & \citeauthor{klassen2022run} &	AI & Speculative Fiction & Non-Experts & 4. Own\\
        2022 & \citeauthor{nunes2022using} &	AI & Speculative Design & Non-Experts & 2. Include\\
        2022 & \citeauthor{stemasov2022ephemeral} &	Wear./Body Enhance. & Speculative Design & Experts & N/A\\
        2023 & \citeauthor{villa2023understanding} &	Wear./Body Enhance. & Scenario-Based & Experts & 1. Consult\\
        2024 & \citeauthor{aljuneidi2023disclose} &	Wear./Body Enhance. & Speculative Design & Combination & 1. Consult\\
        \bottomrule
\end{tabular}
\caption{\label{tab:obj-3}All works in our corpus with the primary objective of using scenarios to provoke ethical reflection and promote critical awareness. Organized chronologically by year, then alphabetically by reference.}
\end{table*}

\begin{table*}[ht!]
    \centering
    \addtolength{\tabcolsep}{-0.18em}
    \rowcolors{2}{LightSteelBlue1!50}{white}
    \begin{tabular}{clllll}
        \toprule
        \multicolumn{6}{c}{\textbf{Objective: Anticipate threats and risks of these technologies}} \\
        \toprule
        Year & Reference & Emerging Technology & Terminology & Scenario Creators & Participation\\
        \midrule
        2019 & \citeauthor{knowles2019scenario} &	IoT & Scenario-Based & Combination & 1. Consult\\
        2020 & \citeauthor{merrill2020security} &	Privacy & Speculative Fiction & Experts & 1. Consult\\
        2021 & \citeauthor{watson2021hci} &	General & Scenario-Based & Experts & 1. Consult\\
        2022 & \citeauthor{tseng2022dark} &	AR/VR & Speculative Design & Non-Experts & 2. Include\\
        2023 & \citeauthor{bonnail2023memory} &	AR/VR & Speculative Design & Experts & 3. Collaborate\\
        2023 & \citeauthor{eghtebas2023co} &	AR/VR & Speculative Design & Combination & 3. Collaborate\\
        2024 & \citeauthor{baldry2024embodied} &	AI & Speculative Design & Combination & 2. Include\\
        2024 & \citeauthor{ehsan2024seamful} &	AI & Scenario-Based & Combination & 3. Collaborate\\
        2024 & \citeauthor{jantunen2024researchers} &	AI & Scenario-Based & Experts & 1. Consult\\
        2024 & \citeauthor{lamparth2024human} &	AI & Scenario-Based & Experts & 1. Consutl\\
        \bottomrule
\end{tabular}
\caption{\label{tab:obj-4}All works in our corpus with the primary objective of using scenarios to anticipate threats and risks of these technologies. Organized chronologically by year, then alphabetically by reference.}
\end{table*}

\begin{table*}[ht!]
    \centering
    \addtolength{\tabcolsep}{-0.18em}
    \rowcolors{2}{LightSteelBlue1!50}{white}
    \begin{tabular}{clllll}
        \toprule
        \multicolumn{6}{c}{\textbf{Objective: Explore perceptions and impact of %sometimes prototype 
        novel technologies}} \\
        \toprule
        Year & Reference & Emerging Technology & Terminology & Scenario Creators & Participation\\
        \midrule
        2017 & \citeauthor{homewood2017turned} &	Healthcare & Speculative Design & Combination & 3. Collaborate\\
        2019 & \citeauthor{stals2019urbanixd} &	Other & %: Smellstick Technology\\
        Speculative Design & Experts & 3. Collaborate\\
        2022 & \citeauthor{kaur2022future} &	AR/VR & Speculative Design & Non-Experts & 4. Own\\
        2023 & \citeauthor{turmo2023futuring} &	AI & Speculative Design & Experts & 1. Consult\\
        2024 & \citeauthor{kim2024respect} &	Other &%: Anti Harrassment Tool\\
        Scenario-Based & Experts & 2. Include\\
        \bottomrule
\end{tabular}
\caption{\label{tab:obj-5}All works in our corpus with the primary objective of using scenarios to explore perceptions and impact of novel technologies. Organized chronologically by year, then alphabetically by reference.}
\end{table*}

\end{document}